\documentclass[a4paper]{coupled2019}

\usepackage{graphicx}
\usepackage{amsmath}
\usepackage{amsfonts}
\usepackage{amssymb}
\usepackage{hyperref}

\title{A Hybrid multiphase model based on lattice Boltzmann method direct simulations}

\author{E. P. Montell\` a$^{*,}$$^{\dag}$, C. Yuan$^a$, B. Chareyre$^{*}$ and A. Gens$^{\dag}$}

\heading{E. P. Montell\`{a}, C. Yuan, B. Chareyre and A. Gens}

\address{$^{*}$  Universit\'e Grenoble Alpes (UGA), 3SR, F-38000 Grenoble, France 
\and
$^{\dag}$Universitat Polit\` {e}cnica de Catalunya (UPC), Campus Nord, 08034 Barcelona, Spain
\and
$^{a}$ University of Calgary, Calgary, AB, Canada T2N 1N4\\
}


\abstract{

By means of the multicomponent Shan-Chen lattice Boltzmann method (LBM), we investigate the multiphase flow through porous media. Despite the excellent accuracy of the LBM, large domains result in unaffordable computational expenses. The Hybrid model developed in this study is based on a pore-network (PN) approach that enhances a decomposition of the granular assembly into small subsets (pore throats). Lattice Boltzmann simulations are performed for each pore throat to determine the hydrodynamic properties (entry capillary pressure, primary drainage curve, liquid morphology, etc) at the microscale. The local properties obtained with LBM are incorporated at the network to solve the larger-scale problem. This strategy leads to a significant decrease of the computation time at the sample-scale compared to a fully resolved method. Fluid morphology and phase distribution are evaluated during the drainage of a small granular assembly using the Hybrid model (PN-LBM). Results are contrasted with those obtained in a fully resolved simulation (LBM). The agreement between the two models illustrates the capability of the Hybrid method, which combines the efficiency of the PN approach and the accuracy of the LBM at the pore scale. 
}

\begin{document}

\section{INTRODUCTION}


Fluid flow has a strong impact on the mechanical behavior of partially saturated granular systems. The hydromechanical response of partially saturated granular material has been studied by means of the Discrete Element Method (DEM) \cite{richefeu2009force,scholtes2012discrete}. Such works were restricted to low liquid content to ensure the pendular regime, where liquid is retained in the form of bridges between particles. When the liquid content is increased, pendular bridges coalesce forming complex liquid morphologies. At this point, the pendular regime is replaced by the funicular regime. Less attention has been devoted to examine the mechanism of wet materials during the funicular regime. Nowadays, three-dimensional images from X-ray tomography are a useful and common tool to characterize the fluid morphology in the porous medium \cite{scheel2008morphological}. Despite these recent advances, very few attempts have been carried out to obtain the details of the liquid distribution within the porous media \cite{melnikov2015grain,yuan2018deformation}.


%

Several modeling techniques can be adopted to reproduce the fluid displacement within the porous media. \textit{Macro-scale models} are based on continuum descriptions of flow. These models incorporate the local fluctuations induced by the microscopic interactions between the fluids ensuring mass conservation. In spite of the low computational cost, macro-scale continuum models rely on empirical relations and information about constitutive properties is required. \textit{Micro-scale continuum models} are fully resolved methods capable to simulate multiphase flow through complex geometries and do not rely on empirical relations. Even though these models have been proven successful to reproduce the physical phenomena at the different scales, they are computationally much more demanding. Lattice Boltzmann method (LBM) \cite{pan2004lattice}, volume of fluid (VOF) method \cite{raeini2012modelling} or Lagrangian mesh-free methods \cite{tartakovsky2007pore} are examples of micro-scale continuum models employed in the analysis of multiphase flow through porous media. Finally, \textit{pore-scale models} manage to reduce the high complexity of porous morphology by a discretization of the pore bodies and pore throats. 


The multiscale method presented in this article is inspired in the pore-scale approach referred as "two-phase pore-scale finite volume-discrete element method" (2PFV-DEM) \cite{yuan2016pore}. In this method, fluid displacement is controlled by the entry capillary pressure ($p^e_c$). When the capllary pressure ($p_c$) exceeds the entry capillary pressure  ($p_c>p^e_c$), the non-wetting phase passes through a narrow throat invading a pore body.  $p^e_c$ can be approximated by the Incircle method \cite{sweijen2016effects}, the Mayer-Stowe-Princen (MS-P) method \cite{yuan2016pore}, or direct fluid simulations. The present Hybrid model follows the pore-scale decomposition scheme proposed by \cite{chareyre2012pore} and relies on LBM simulations to determine the main hydrostatic properties ($p^e_c$, liquid morphology, capillary pressure - saturation curves).


\section{NUMERICAL METHODS}

\subsection{Lattice Boltzmann Method}

A multicomponent multiphase Shan-Chen LB model \cite{shan1993lattice} is implemented using the open source library of Palabos. The motion of a fluid is described by the lattice Boltzmann equation:

\begin{equation}
\label{LBM_equation}
f_k^\omega (\boldsymbol {x_k + e_k}\Delta t, t+ \Delta t)-f_k^\omega (\boldsymbol {x_k}, t)=\dfrac{-\Delta t}{\tau ^\omega} (f_k^\omega (\boldsymbol {x_k}, t)-f_k^{\omega,eq} (\boldsymbol {x_k}, t))
\end{equation}

where the superscript $\omega$ denotes the $\omega$th fluid, $\tau^\omega$ is the rate of relaxation towards local equilibrium,  $f_k^{\omega,eq}$ is the equilibrium distribution function, $\Delta t$ is the time increment, $\boldsymbol {e_k}$ are the discrete velocities which depend on the velocity model, in this work, D3Q19 (three-dimensional space and 19 velocities) model is adopted, and $k$ varies from 0 to $Q-1$ representing the directions in the lattice. The left-hand side of  Eq. \ref{LBM_equation} corresponds to the streaming step, which describes the motion of the fluid particles between nodes, whereas the right-hand side stands for the collision step based on the Bhatnagar-Gross-Krook (BGK) approximation. The collision operator represents the evolution of the system towards the equilibrium.

The macroscopic density and momentum variables are recovered from the distribution functions:

\begin{equation}\label{eqn:MOMENT_0}
\rho^{\omega} = \sum_k f_k^{\omega}
\end{equation}

\begin{equation}\label{eqn:MOMENT_1}
\rho^{\omega} \mathbf{u^{\omega}}= \sum_k f_k^{\omega} \mathbf{e_k}
\end{equation}


In order to ensure a system of immisible fluids, a repulsive force between the fluid phases is introduced. According to \cite{shan1993lattice}, the non-local force responsible for the fluid-fluid interaction can be expressed as: 

\begin{equation}\label{eqn:interparticle_force_LBM}
\mathbf{F_\omega (x)} = -\Psi(\mathbf {x}) \sum_{\bar{\omega}} G_{\sigma \bar{\omega}}\sum_k \Psi_{\bar{k}} \mathbf{(x+e_k) e_k}
\end{equation}

where $\Psi_k$ is the interparticle potential that induces phase separation and $G_{\omega \bar{\omega}}$ is the interaction strength between components $\omega$, $\bar{\omega}$.

Finally, the non-ideal equation of state can be expressed as:

\begin{equation}\label{eqn:interparticle_force_LBM}
p = c_s^2 ( \rho_\omega + \rho_{\bar{\omega}}) +  c_s^2 G \rho_\omega  \rho_{\bar{\omega}}
\end{equation}

where $ \rho_\omega$ and $\rho_{\bar{\omega}}$ are the densities of the fluids at a certain position and $c_s$ is the speed of sound.


%
%
%
%
%
%
%
%
%
%

\subsection{PN-LBM Hybrid model}

\subsubsection{Pore-space decomposition }

\begin{figure}[h]
\centering
\includegraphics[width=14cm]{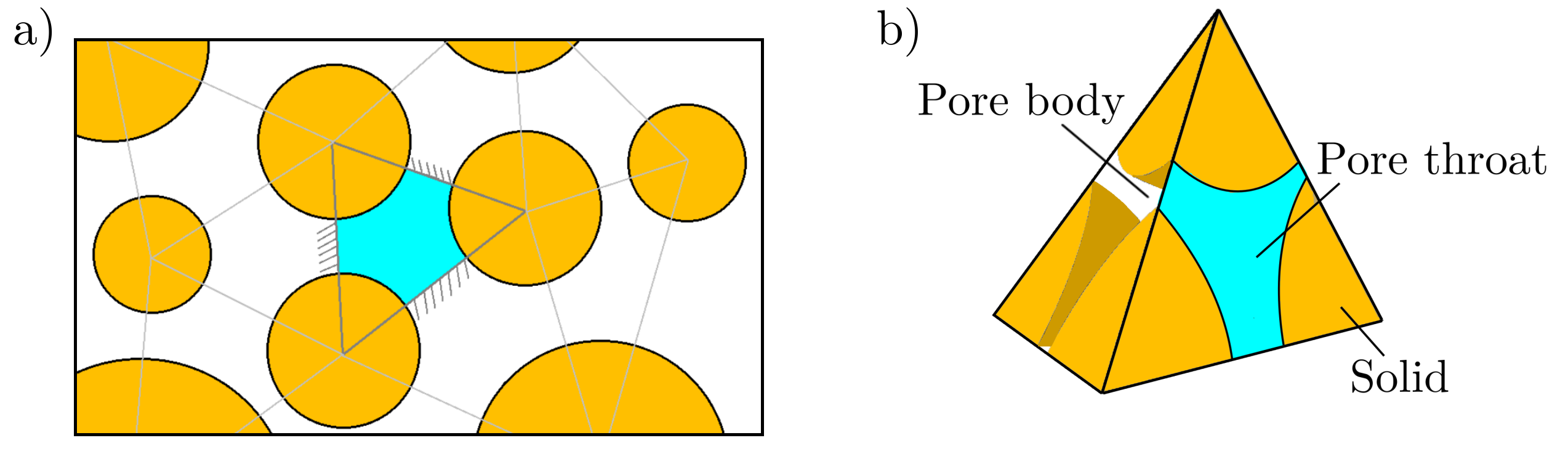}
\caption{\label{fig:decomposition_scheme} Adjacent tetrahedron in the regular triangulation of a granular assembly in two dimensions (a) and three dimensions (b).}
\end{figure}

The pore-scale network employed in the present Hybrid method is based on the 2PFV-DEM scheme developed by \cite{yuan2016pore}, which combines a three-dimensional triangulation method and DEM to model the hydro-mechanical behavior of unsaturated deformable granular materials. The triangulation of the pore space leads to a tetrahedral mesh whose vertices coincide with the centers of the spheres. Each tetrahedral element defines a pore body and four pore throats (see tetrahedral facets in figure \ref{fig:decomposition_scheme}a). A 2D schematic diagram is included in figure \ref{fig:decomposition_scheme}b to illustrate the discretization of the void space. This technique is particularly convenient to describe the topology of the pore throats.



%




\subsubsection{Fluid displacement}
\label{subsubsubsec:fluid_di}

The fluid front displacement and the possible combinations of fluid in the porous media are mainly controlled by the entry capillary pressure. Following the work of \cite{Chandler1982}, the displacement of the non-wetting phase is directly link to the capillary pressure. The interface curvature increases as the capillary pressure builds up following Laplace-Young equation. When the local capillary pressure is larger than the entry capillary pressure, the interface front becomes unstable and the non-wetting phase passes through the pore throat invading the pore body. According to \cite{orr1977three}, the analysis of the capillary pressure and meniscus shape requires solving the Laplace-Young equation, a nonlinear, second-order, partial differential equation which, can be numerically solved for very few 3-spheres configurations. Due to this limitation, it seems reasonable to consider other approaches that can predict the entry capillary pressure and liquid morphology.

\begin{figure}[h]
\centering
\includegraphics[width=15cm]{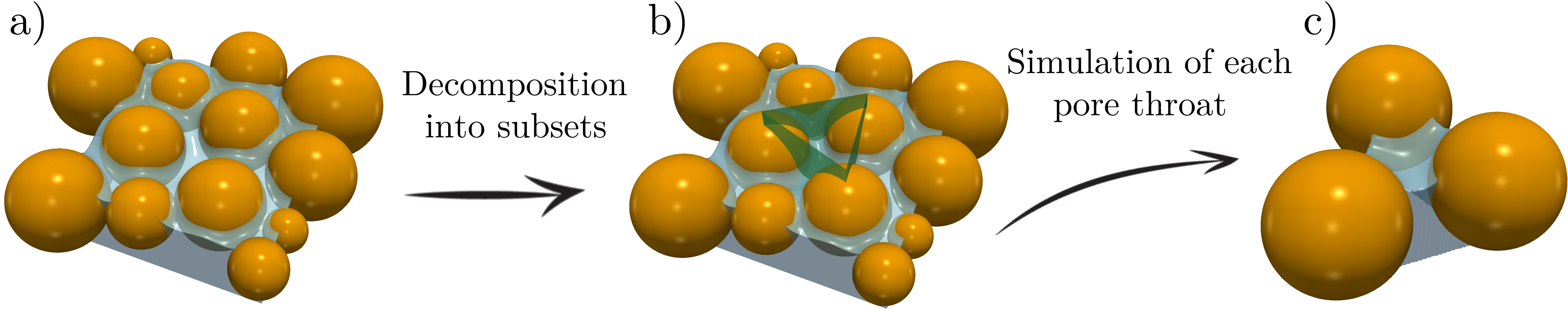}
\caption{\label{fig:decomposition_scheme_Hybrid} Decomposition of the granular assembly (a) into small subsets (b). Each subset is made up of 3 spheres (c).}
\end{figure}

The pore space is decomposed into a series of throat-domains following a regular triangulation (see figure \ref{fig:decomposition_scheme_Hybrid}). The decomposition leads to a list of pore throats that are solved independently. LBM simulation are performed for each pore throat to estimate the primary drainage curve, the entry capillary pressure and liquid morphology. The computation domain is a triangular-shaped prism defined by three solid walls orthogonal to the pore throat. Each of these solid boundaries passes through two of the spheres centers (two vertex of the triangle defined by the 3 spheres centers). The reader is directed to figure \ref{fig:decomposition_scheme_Hybrid} for a more comprehensive review . The domain is enclosed by two triangles at the top and bottom of the prism representing the inlet and outlet sections of the LBM simulations.


\begin{figure}[h]
   \centering
\includegraphics[width=10cm]{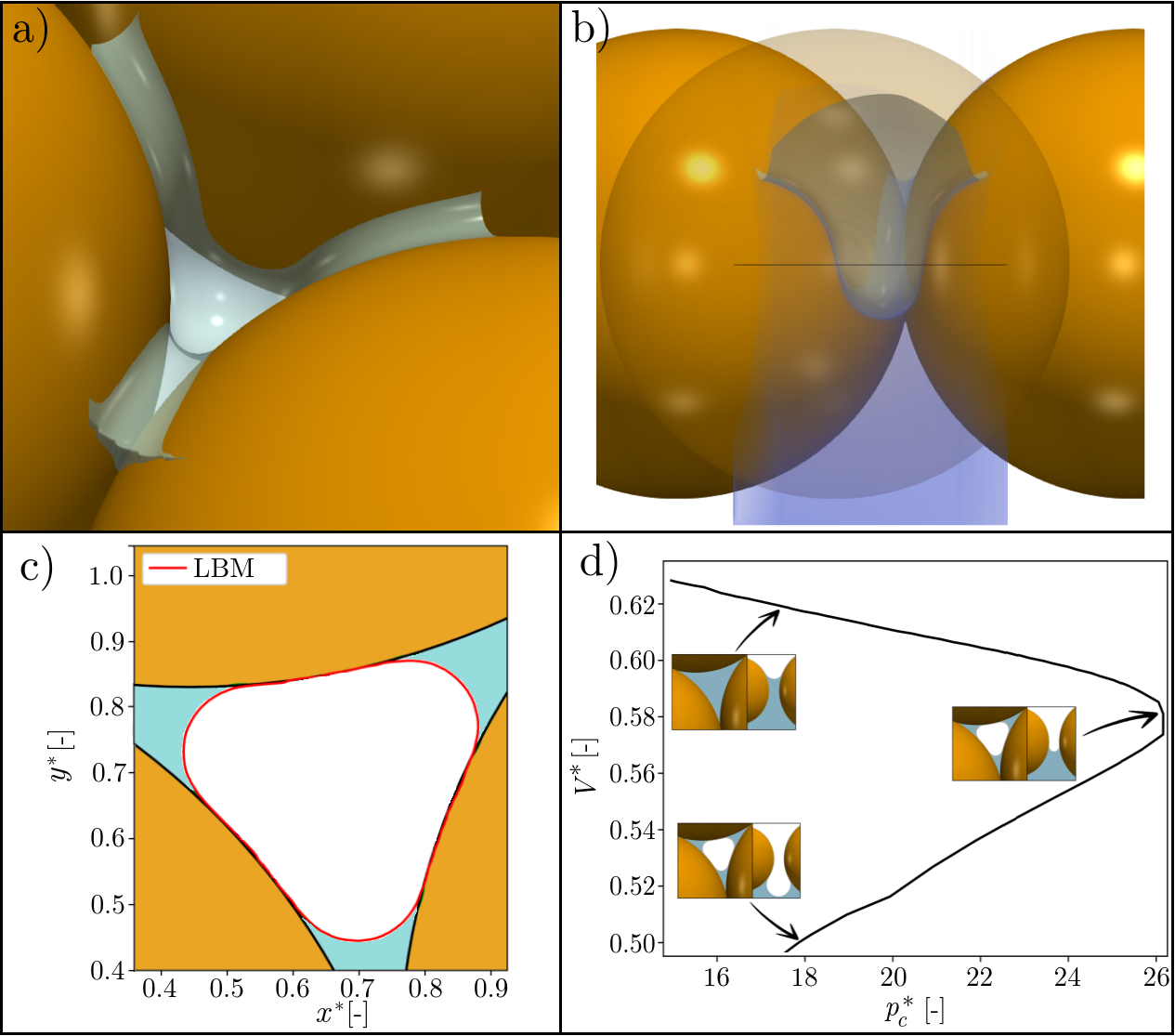}   
    \caption{a) Fluid-fluid interface deforming due to increments of capillary pressure. b) Vertical view of the interface at the moment of the pore throat invasion. c) Wetting phase is accumulated in the corners while in the center non-wetting phase fills the pore throat. d) Capillary pressure - Saturation curve for the pore thorat. Three snapshots displaying vertical and horizontal slices of the simulation evidence the liquid morphology before, during and after the pore invasion.}
    \label{fig:3D_combo}
\end{figure}

Figure \ref{fig:3D_combo}d shows the evolution of a typical subset. In this case, the pore throat is formed by three equal-sized spheres in contact. The density of the nodes located in the inlet and outlet sections is adjusted to gradually increase the capillary pressure. Such increment induces the displacement of the interface (see figures \ref{fig:decomposition_scheme_Hybrid}c and figure \ref{fig:3D_combo}).


The evolution of normalized capillary pressure ($p_{c}^*=\dfrac{2 p_{c}R}{\gamma}$, where $\gamma$ is the surface tension, $R$ is the radius of the spheres and $p_{c}$ is the capillary pressure) during the invasion is related to the change of volume of the wetting phase in figure \ref{fig:3D_combo}d. Dimensionless volume is defined as: $V^*=\dfrac{V(t)}{\dfrac{4}{3} \pi R^3}$, where $V(t)$ is the volume at time $t$ .

The entry capillary pressure ($p^{e}_c$) is defined as the maximum value reached by $p_c$ during the invasion process. When the capillary pressure reaches the entry capillary pressure ($p_c=p^e_c$), the non-wetting phase penetrates into the pore body (see figure \ref{fig:3D_combo}). Immediately after the invasion, the interface meniscus expands leading to a reduction of the capillary pressure. This procedure is repeated for all the pore throats of the sample. Thus, $p^e_c$ is determined for all the subdomains and incorporated into the global problem handled by the pore-network.

\section{RESULTS}


\subsection{Flow through a 40 spheres  granular assembly}

This section presents the results of a drainage process in a 40 sphere granular assembly.  A fully resolved LBM simulation has been performed and assumed as reference data. Then, results are compared with the Hybrid model.

\subsubsection{Numerical setup}

A random sphere pack is created by the open-source code YADE \cite{vsmilauer2010yade}. A cubic box of 10 mm x 10 mm x 10 mm is defined in which 40 polydisperse spheres are packed. The mean sphere radius is 1.26 mm. The following assumptions are made during the numerical simulations:

\begin{itemize}	
 \item[$-$] Negligible gravitational forces.
 \item[$-$] Static solid skeleton.
 \item[$-$] Drainage is evaluated under quasistatic flow.
 \item[$-$] Perfect wetting of the solid by the wetting phase.
 \item[$-$] Disconnected regions remain saturated by a fixed amount of wetting phase.
\end{itemize}

\begin{figure}[h]
\centering
\includegraphics[width=7cm]{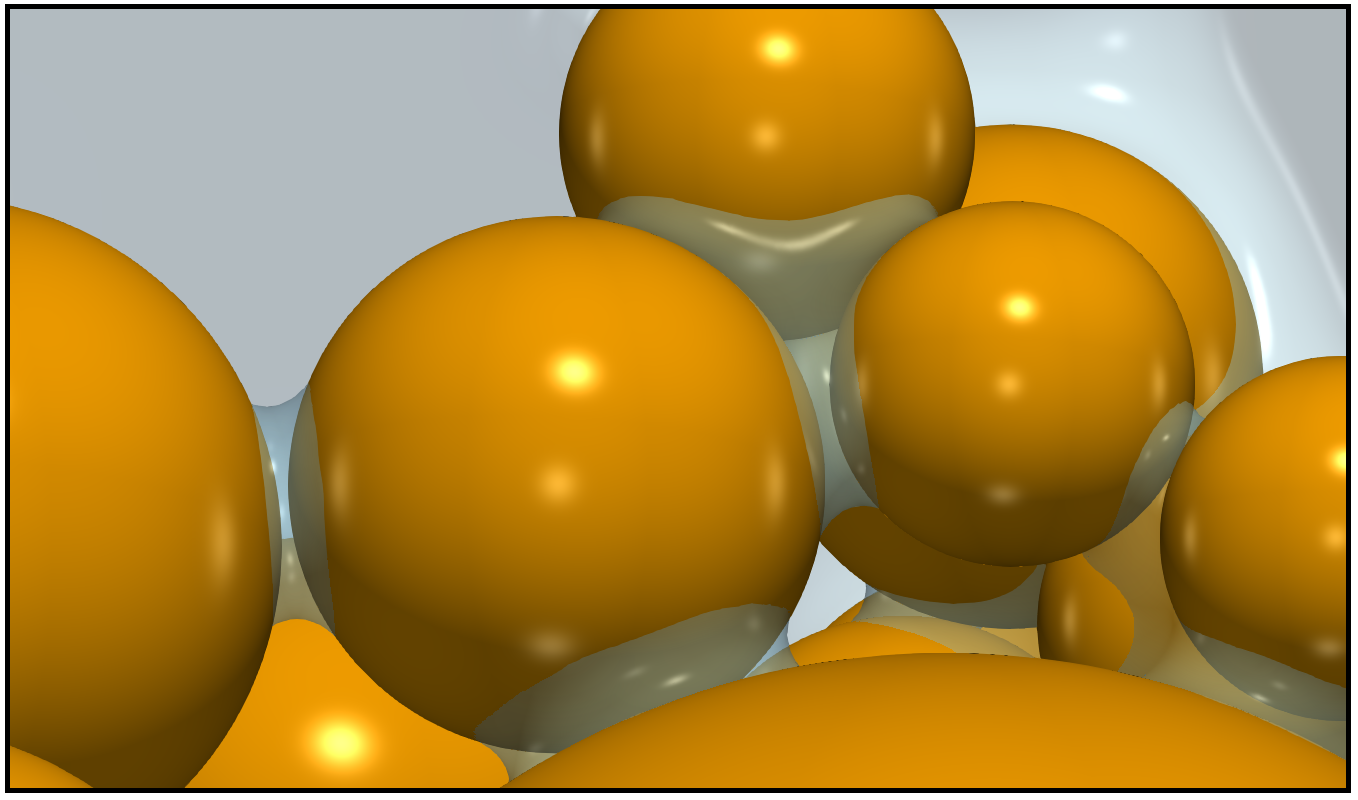}
\caption{\label{fig:trimer_bridge_LBM_40_spheres} Distribution of the wetting phase after the drainage simulation of a 40 sphere packing. The wetting phase is trapped in some areas of the sample as pendular bridges or liquid clusters.}
\end{figure}

\begin{figure}[h]
\centering
\includegraphics[width=15cm]{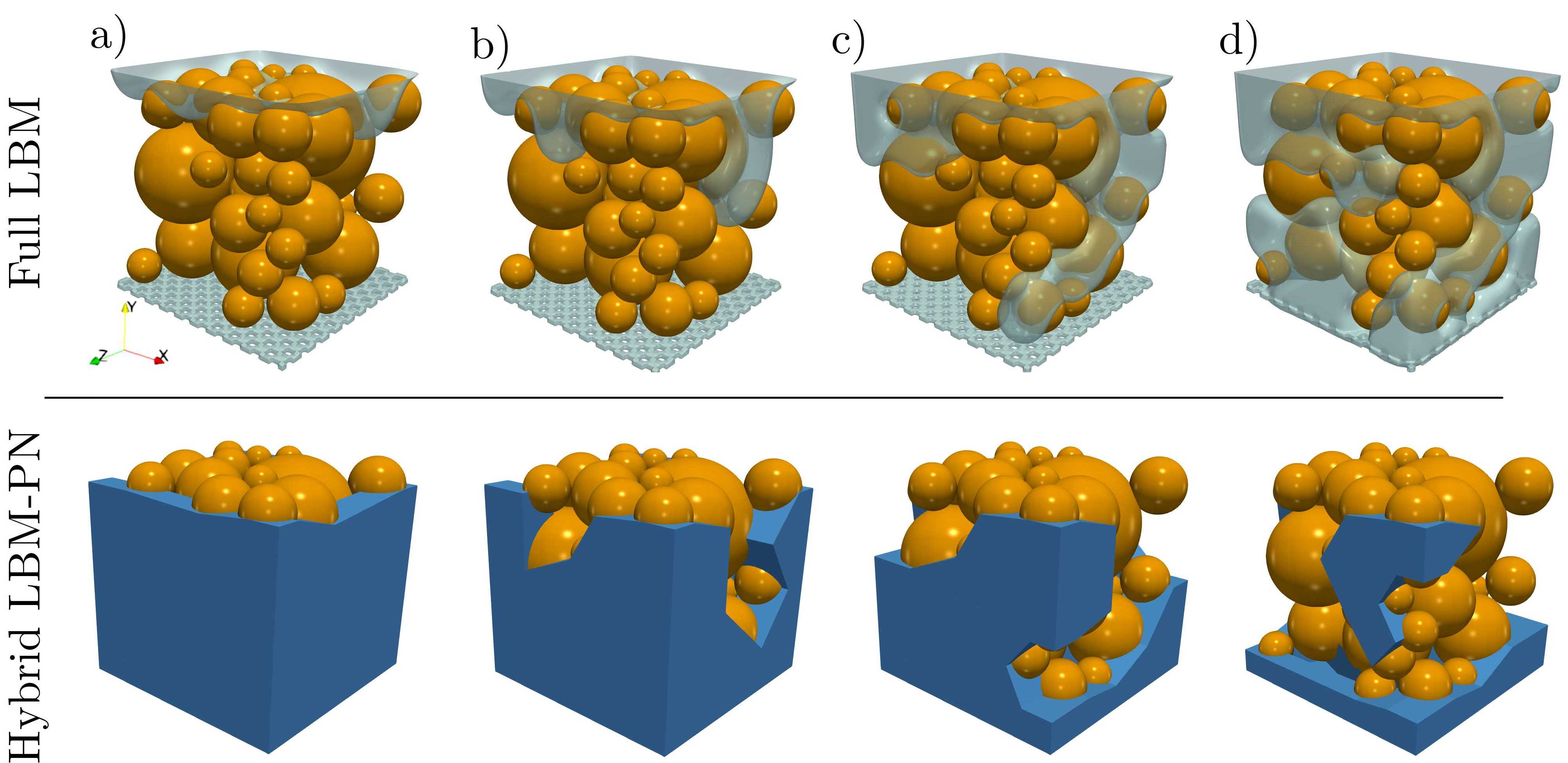}
\caption{\label{fig:Path_tot_FRONT} Comparison of the invasion path between the full LBM simulation (translucent interface) and Hybrid model (wetting phase depicted in blue).}
\end{figure}

Initially, the granular assembly is completely saturated (see figure \ref{fig:Path_tot_FRONT}a - Full LBM). A porous membrane is located at the bottom of the sample in order to prevent the non-wetting phase to reach the outlet and ensure a complete drainage. The displacement of the interface is conducted by a fixed flow controlled by a pressure difference between the non-wetting reservoir (top of the sample) and the wetting reservoir (bottom of the granular assembly). The interface flows towards the bottom driven by the pressure gradient invading the largest pores at first, then, the non-wetting phase progressively occupies the pore spaces of the sample. Capillary pressure is increased and recorded until all the nodes above the porous plate are filled with the non-wetting phase (see figure \ref{fig:Path_tot_FRONT}d- Full LBM). At this moment, the remaining wetting phase is trapped in the granular assembly in form of liquid clusters. Figure \ref{fig:trimer_bridge_LBM_40_spheres} evidences the presence of pendular bridges and trimers (liquid cluster formed between three grains) in the sphere packing after the drainage.

\subsubsection{Invasion paths}

The aim of the present section is to characterize the invasion path of the non-wetting phase through the porous media. Figure \ref{fig:Path_tot_FRONT} captures the non-wetting pathways for different time steps for each method.  The full LBM interface is shown as translucid blue isosurface and the wetting phase is depicted in blue for the Hybrid model in figure \ref{fig:Path_tot_FRONT}. We remark the ability of Hybrid model to predict the first preferential path (identical intrusion are observed in figures \ref{fig:Path_tot_FRONT}b and \ref{fig:Path_tot_FRONT}c). As the drainage proceeds, the non-wetting phase keeps occupying the sample leaving behind some disconnected liquid clusters (see figure \ref{fig:Path_tot_FRONT}d). Even though wetting phase trapped in the assembly is well represented with the Hybrid method, figure \ref{fig:Path_tot_FRONT}d suggests that there is a certain phase mismatch between the models in some regions.
%
%

It is worth mentioning that the CPU time for a complete drainage corresponds to   29.6 days for the full LBM. On the other hand, by using the Hybrid model, the computational cost is reduced down to 11.2 days.

\section{CONCLUSIONS}

Compared to a fully resolved method (in this work: LBM), the Hybrid model presented in this article provides more efficient results in terms of computation time. Results show that Hybrid method is able to mimic the interface displacement during the drainage of pore space with excellent accuracy. 


This technique would probably have a stronger impact on a lager problem with thousands of pore throats. In parallel simulations with relatively small computational domains, the execution speed is multiplied by a factor two doubling the number of cores. Large-scale problems that require the exploitation of massively parallel systems are frequently accompanied by a drop in efficiency. In the Hybrid method, however, the scalability is optimal. Due to the fact that each pore throat (small domains) are assignated to a single core, efficiency does not depend on the size of the granular assembly. Besides, some regions of the sample can be excluded from the simulation (empty pores and isolated cluster with no flux).

\bibliographystyle{ieeetr}
\bibliography{biblioEDU}

\begin{thebibliography}{10}

\bibitem{richefeu2009force}
V.~Richefeu, M.~S. El~Youssoufi, E.~Az{\'e}ma, and F.~Radjai, ``Force
  transmission in dry and wet granular media,'' {\em Powder Technology},
  vol.~190, no.~1-2, pp.~258--263, 2009.

\bibitem{scholtes2012discrete}
L.~Scholt{\`e}s, B.~C.~F. Nicot, and F.~Darve, ``Discrete modelling of
  capillary mechanisms in multi-phase granular media,'' {\em arXiv preprint
  arXiv:1203.1234}, 2012.

\bibitem{scheel2008morphological}
M.~Scheel, R.~Seemann, M.~Brinkmann, M.~Di~Michiel, A.~Sheppard,
  B.~Breidenbach, and S.~Herminghaus, ``Morphological clues to wet granular
  pile stability,'' {\em Nature Materials}, vol.~7, no.~3, p.~189, 2008.

\bibitem{melnikov2015grain}
K.~Melnikov, R.~Mani, F.~K. Wittel, M.~Thielmann, and H.~J. Herrmann,
  ``Grain-scale modeling of arbitrary fluid saturation in random packings,''
  {\em Physical Review E}, vol.~92, no.~2, p.~022206, 2015.

\bibitem{yuan2018deformation}
C.~Yuan, B.~Chareyre, and F.~Darve, ``Deformation and stresses upon drainage of
  an idealized granular material,'' {\em Acta Geotechnica}, vol.~13, no.~4,
  pp.~961--972, 2018.

\bibitem{pan2004lattice}
C.~Pan, M.~Hilpert, and C.~Miller, ``Lattice-boltzmann simulation of two-phase
  flow in porous media,'' {\em Water Resources Research}, vol.~40, no.~1, 2004.

\bibitem{raeini2012modelling}
A.~Q. Raeini, M.~J. Blunt, and B.~Bijeljic, ``Modelling two-phase flow in
  porous media at the pore scale using the volume-of-fluid method,'' {\em
  Journal of Computational Physics}, vol.~231, no.~17, pp.~5653--5668, 2012.

\bibitem{tartakovsky2007pore}
A.~M. Tartakovsky, A.~L. Ward, and P.~Meakin, ``Pore-scale simulations of
  drainage of heterogeneous and anisotropic porous media,'' {\em Physics of
  Fluids}, vol.~19, no.~10, p.~103301, 2007.

\bibitem{yuan2016pore}
C.~Yuan, B.~Chareyre, and F.~Darve, ``Pore-scale simulations of drainage in
  granular materials: finite size effects and the representative elementary
  volume,'' {\em Advances in water resources}, vol.~95, pp.~109--124, 2016.

\bibitem{sweijen2016effects}
T.~Sweijen, E.~Nikooee, S.~M. Hassanizadeh, and B.~Chareyre, ``The effects of
  swelling and porosity change on capillarity: Dem coupled with a pore-unit
  assembly method,'' {\em Transport in porous media}, vol.~113, no.~1,
  pp.~207--226, 2016.

\bibitem{chareyre2012pore}
B.~Chareyre, A.~Cortis, E.~Catalano, and E.~Barth{\'e}lemy, ``Pore-scale
  modeling of viscous flow and induced forces in dense sphere packings,'' {\em
  Transport in porous media}, vol.~94, no.~2, pp.~595--615, 2012.

\bibitem{shan1993lattice}
X.~Shan and H.~Chen, ``Lattice boltzmann model for simulating flows with
  multiple phases and components,'' {\em Physical Review E}, vol.~47, no.~3,
  p.~1815, 1993.

\bibitem{Chandler1982}
R.~Chandler, J.~Koplik, K.~Lerman, and J.~F. Willemsen, ``Capillary
  displacement and percolation in porous media,'' {\em Journal of Fluid
  Mechanics}, vol.~119, pp.~249--267, 6 1982.

\bibitem{orr1977three}
F.~Orr~Jr, R.~Brown, and L.~Scriven, ``Three-dimensional menisci: Numerical
  simulation by finite elements,'' {\em Journal of Colloid and Interface
  Science}, vol.~60, no.~1, pp.~137--147, 1977.

\bibitem{vsmilauer2010yade}
V.~{\v{S}}milauer and B.~Chareyre, ``Yade dem formulation,'' {\em Yade
  Documentation}, 2010.

\end{thebibliography}


\end{document}